\documentclass[11pt]{article}

\usepackage[preprint]{acl}

\usepackage{acl} 
\usepackage{times}
\usepackage{latexsym}
\usepackage{inconsolata} 
\usepackage{microtype}
\usepackage{graphicx}
\usepackage{booktabs}
\usepackage{multirow}
\usepackage{makecell}
\usepackage{amsmath, amssymb, amsthm, mathtools}
\usepackage{colortbl}
\usepackage{xcolor}
\usepackage{enumitem}
\usepackage{array}
\usepackage{pifont}
\usepackage{url}
\usepackage{listings}
\usepackage{algorithm}
\usepackage{algpseudocode}
\usepackage{adjustbox}
\usepackage{tikz}
\usepackage{caption}
\usepackage{subcaption}

\newcommand{\model}{\textsc{DSL-R1}}

\lstdefinestyle{dsl}{
  basicstyle=\ttfamily\small,
  keywordstyle=\color{blue}\bfseries,
  commentstyle=\color{teal!60!black}\itshape,
  stringstyle=\color{orange!70!black},
  showstringspaces=false,
  frame=single,
  breaklines=true
}
\lstdefinelanguage{DSL}{
  morekeywords={SELECT,FROM,WHERE,AND,OR,ORDER,BY,LIMIT,ASC,DESC,EMBED,TOPK,VECTOR_SIM,MATCH},
  sensitive=true,
  morecomment=[l]{\#},
  morestring=[b]"
}

\usepackage{times}
\usepackage{latexsym}

\usepackage[T1]{fontenc}

\usepackage[utf8]{inputenc}

\usepackage{microtype}

\usepackage{inconsolata}

\usepackage{graphicx}
\usepackage{listings}
\usepackage{xcolor}

\title{DSL-R1: From SQL to DSL for Training Retrieval Agents across Structured and Unstructured Data with Reinforcement Learning}

\author{
  \textbf{Yunhai Hu\textsuperscript{1*}},\thanks{~Corresponding Author.}
  \textbf{Junwei Zhou\textsuperscript{2*}},
  \textbf{Yumo Cao\textsuperscript{3}},
  \textbf{Yitao Long\textsuperscript{1}},
\\
  \textbf{Yiwei Xu\textsuperscript{3}},
  \textbf{Qiyi Jiang\textsuperscript{3}},
  \textbf{Weiyao Wang\textsuperscript{3}},
  \textbf{Xiaoyu Cao \textsuperscript{3}},
\\
  \textbf{Zhen Sun\textsuperscript{3}},
  \textbf{Yiran Zou\textsuperscript{3}},
  \textbf{Nan Du\textsuperscript{2*}},
\\
\\
  \textsuperscript{1}New York University,
  \textsuperscript{2}Matter Innovation Inc.,
  \textsuperscript{3}Thin Red Line
\\
  \small{
    \href{mailto:yh5961@nyu.edu}{yh5961@nyu.edu},
    \href{mailto: junweizhou@matter.ai}{ junweizhou@matter.ai}, \href{mailto: frankdu@matter.ai}{ frankdu@matter.ai}
  }
}
\begin{document}
\maketitle
\begin{abstract}
Effective retrieval in complex domains requires bridging the gap between structured metadata and unstructured content. Existing systems typically isolate these capabilities, relying on either symbolic filtering or vector similarity, failing to capture their interplay. In this work, we propose DSL-R1, a unified framework that synergizes logical reasoning with semantic matching via a novel Domain-Specific Language (DSL). By embedding vector primitives within SQL-style operators, our approach leverages the complementary strengths of symbolic precision and semantic coverage. We further introduce a reinforcement learning mechanism where rule-based execution feedback and retrieval quality rewards jointly optimize the DSL generation, balancing structural correctness and semantic alignment. Evaluations on a large-scale industrial email benchmark demonstrate that DSL-R1 achieves a +12.3\% improvement in Hit@1/3, consistently outperforming decoupled baselines and establishing a robust paradigm for hybrid retrieval.
\end{abstract}

\section{Introduction}

The explosive growth of  data presents pressing challenges for retrieval systems. Symbolic query languages such as SQL~\citep{zhong2017seq2sql} or SPARQL~\citep{berant2013semantic} provide strong interpretability and precise logical control but struggle with unstructured modalities such as free-form text or document content.  
Conversely, vector-based models (e.g., CLIP~\citep{radford2021clip}, BLIP-2~\citep{li2023blip2}) excel at semantic matching but lack the ability to enforce logical constraints and offer explainability. Bridging these two paradigms is critical for building retrieval systems that are both robust and interpretable.

To this end, we propose a DSL-driven retrieval framework that integrates SQL operators with vector search under a unified syntax. This design enables compositional queries that capture both logical constraints and semantic intents within a single, language-like interface.  
A key challenge lies in generating and executing DSL queries reliably under ambiguous or noisy user inputs. We address this by introducing a reinforcement learning framework that jointly optimizes correctness and retrieval utility through rule-based reward functions, enhancing both execution robustness and semantic coverage.

We evaluate our framework on a large-scale retrieval dataset constructed from business emails, which combines structured metadata with unstructured text and attachments. To demonstrate cross-domain robustness, we further validate our approach on the ArxivQA benchmark. Experiments using GRPO~\citep{grpo2024} and DAPO~\citep{dapo2025} optimization algorithms show substantial accuracy gains over symbolic-only and vector-only baselines, with DAPO offering the best trade-off between stability and efficiency.

Our main contributions are summarized as follows:
\textbf{Large-scale hybrid retrieval benchmarks:} we construct a challenging dataset from business emails integrating structured metadata with unstructured textual and visual content.
\textbf{A DSL-agent reinforcement learning framework:} a unified DSL combining SQL-style logical operators with vector-based retrieval, optimized through RL signals using LLM-simulated feedback. Experiments demonstrate that our approach significantly outperforms baselines, with DAPO providing the best trade-off between stability and efficiency.

\begin{figure*}[!t]
  \centering
  \includegraphics[width=0.9\linewidth]{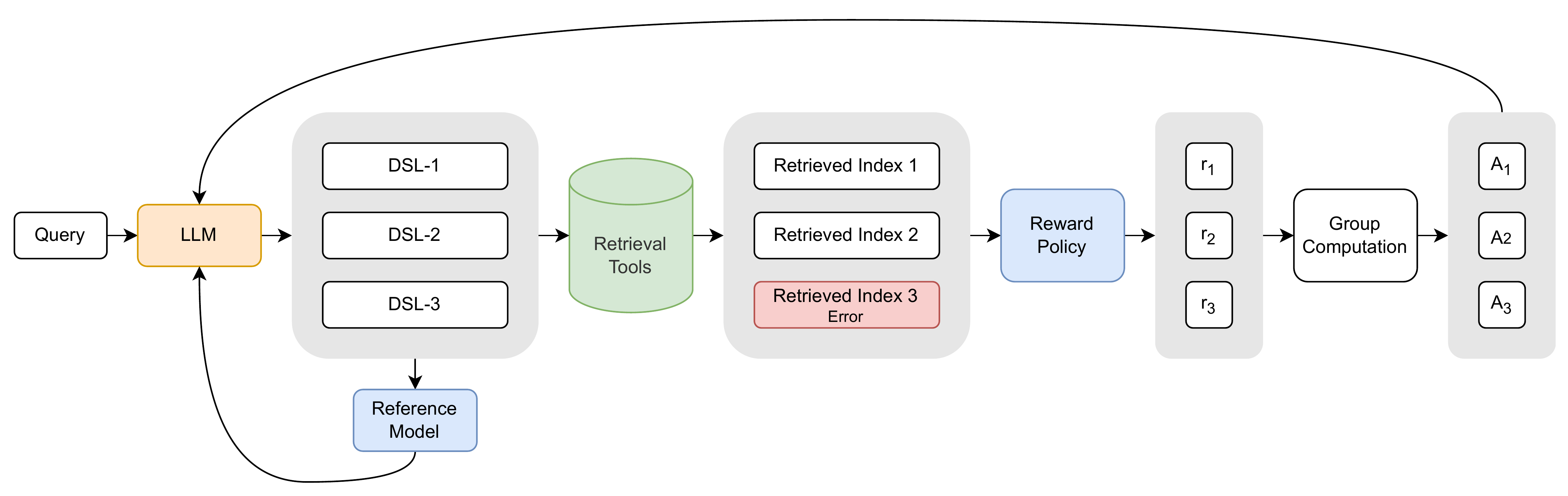}
  \caption{Overview of the reinforcement learning framework for DSL-based  retrieval}
  \label{fig:rl-overview}
\end{figure*}

\section{Related Work}
\subsection{Retrieval}
Retrieval methods have revolutionized the processing of unstructured data by mapping inputs into a semantic space. Early works such as DPR~\citep{karpukhin2020dense} and ANCE~\citep{xiongapproximate} demonstrated the effectiveness of dual-encoder architectures for text retrieval.

This paradigm has been successfully extended to multimodal contexts, where models like CLIP~\citep{radford2021clip}, ALIGN~\citep{jia2021align}, and BLIP-2~\citep{li2023blip2} capture semantic alignment across text, image, and video modalities. 
In contrast, symbolic retrieval methods like SQL~\citep{zhong2017seq2sql} and SPARQL~\citep{berant2013semantic} offer structural precision and interpretability but are less effective for unstructured or ambiguous inputs.  
Recent hybrid methods, including ColPali~\citep{colpali2024}, and ColQwen~\citep{colqwen2024}, enhance dense retrieval with contextualized representations, while fusion-based systems such as FiD~\citep{izacard2021fid}, and InstructBLIP~\citep{dai2023instructblip} explore  reasoning by integrating retrieval and generation.  
However, most existing works operate at the representation level, lacking an explicit symbolic layer for compositional reasoning. We address this by introducing a trainable DSL that unifies symbolic operators with vector retrieval, optimized via reinforcement learning to ensure both logical correctness and semantic quality.

\subsection{DSLs in AI Systems}

DSLs have been widely applied in program synthesis~\citep{balog2017deepcoder, nye2021show, chen2021program}, data analysis~\citep{wang2023dslsql, zhang2024datadsl}, and reasoning with language models~\citep{jiang2023program, wei2022chain, gupta2023programmatic}.  
DSLs enable structured expressivity and verifiable reasoning but traditionally lack adaptability without external optimization or learning signals.  
Recent advances have explored combining DSLs with reinforcement learning or large language models to improve control and generalization~\citep{xu2024rlcode, yao2023react, gou2024llmcompiler}.  
Our framework continues this direction by treating the DSL not as a static interface but as a learnable policy space, jointly optimized through rule-based rewards and feedback to achieve semantic alignment and robust execution in  retrieval contexts.

\section{Method}

\subsection{Overview}
Our framework consists of three main \emph{components}: 
(i) a \emph{DSL-based  retrieval agent}, which translates natural-language inputs into executable programs that combine structured operators with vector-based similarity search; 
(ii) a \emph{data generation component}, which constructs  retrieval corpora containing both structured and unstructured elements, and produces paired queries, programs, and gold targets to support supervised and evaluative training; 
and (iii) a \emph{reinforcement learning mechanism}, which optimizes the agent policy by leveraging execution-derived rewards and preference-based regularization. Together, these components ensure both interpretability and adaptability, enabling effective retrieval across heterogeneous data~sources.

\subsection{DSL Agent Design}
The core of our system is a \emph{DSL-based  retrieval agent} that integrates symbolic and vector-based operators within a unified execution framework. 
Specifically, the DSL program is composed of two parts: (i) an \emph{SQL-like clause} that handles structured conditions over metadata, and (ii) a \emph{vector search query list} that retrieves candidate items from unstructured modalities. Case samples are included in Appendix~\ref{cases}.

Given a natural-language query $x$, the agent first generates a DSL program $y$ containing both structured filters and vector queries. 
The vector search module computes nearest neighbors in the embedding space and returns a set of candidate primary keys $\{id_1, \ldots, id_k\}$. 
These candidate identifiers are then injected into the SQL execution engine as constraints, where structured predicates are applied. 
The final result set is obtained by merging the outcomes of the SQL filter with the candidate list returned by the vector search. 

This design allows the agent to exploit the precision of symbolic filtering while leveraging the semantic coverage of vector similarity, ensuring both interpretability and robustness in  retrieval tasks.

\subsection{Data Preparation}

Our corpus $\mathcal{D}$ consists of email records $d=(m,u)$, where $m$ denotes structured metadata (sender, date, category) and $u$ represents unstructured content (subject, body, attachments). For each record, we randomly select $1$–$3$ metadata attributes to form a structured filter $F_s$, retrieving a candidate set $\mathcal{C}_s = \{ d' \in \mathcal{D} \mid m'(a_i)=m(a_i)\}$. A gold record $\hat{d}$ is then sampled from $\mathcal{C}_s$, and several semantic cues from its content $u$ are used to construct a query embedding $q$. A vector search refines the candidate set to $\mathcal{C}_{su} = \{ d' \in \mathcal{C}_s \mid \cos(q,u') \ge \tau \}$, where $\tau$ is a similarity threshold. The structured filter $F_s$ and semantic query $q$ are combined into an executable DSL program $y$ that retrieves $\hat{d}$, while a corresponding natural-language query $x$ is generated from the same attributes, forming a supervised triplet $(x, y, \hat{d})$ that couples symbolic constraints with semantic intent and provides supervision. More details are provided in Appendix ~\ref{datasets}.

\subsection{Reinforcement Learning}
As shown in Figure~\ref{fig:rl-overview}, given a natural language query $x$, the policy model generates multiple DSL candidates $\{y_i\}_{i=1}^G$. 
Each candidate is executed with retrieval tools; valid results are scored by the reward function, and errors receive zero or negative scores. 
The group of rewards $\{r_i\}$ is then aggregated to compute group-relative advantages $\{A_i\}$, which guide policy updates.

We adopt two algorithms: GRPO~\citep{grpo2024} and its improved variant DAPO~\citep{dapo2025}.
GRPO enables efficient policy optimization without a value model, while DAPO further stabilizes training through dual-anchor regularization and adaptive reward scaling.

The simplified GRPO objective is:
{\small
\begin{align}
\mathcal{L}_{\text{GRPO}}(\theta) 
= - \frac{1}{G} \sum_{i=1}^G 
& \;\min\!\big(r_i A_i,\; \mathrm{clip}(r_i) A_i\big) \nonumber \\
& + \;\beta D_{\mathrm{KL}}\!\left(\pi_\theta \,\|\, \pi_{\text{ref}}\right),
\end{align}}
where $r_i = \tfrac{\pi_\theta(y_i|x)}{\pi_{\text{old}}(y_i|x)}$ is the importance ratio, $\pi_\theta$ denotes the current policy parameterized by $\theta$,
$\pi_{\text{old}}$ is the behavior policy used to generate the sampled trajectories,
$\pi_{\text{ref}}$ is a fixed reference policy used to regularize the optimization via the KL term,
and $A_i$ is the group-relative advantage.

DAPO further improves GRPO by normalizing advantages at the token level and using asymmetric clipping bounds. Its simplified loss is:
{\small
\begin{align}
\mathcal{L}_{\text{DAPO}}(\theta) 
= - & 2\frac{1}{N} \sum_{i=1}^G \sum_{t=1}^{|y_i|}
 \;\min\!\big(r_{i,t} A_{i,t}, \nonumber \\
& \;\;\;\mathrm{clip}(r_{i,t}, 1-\epsilon_l, 1+\epsilon_h)\, A_{i,t}\big),
\end{align}
}
where $N = \sum_i |y_i|$ is the total number of tokens,  
$r_{i,t} = \tfrac{\pi_\theta(y_{i,t}\mid x,y_{i,<t})}{\pi_{\text{old}}(y_{i,t}\mid x,y_{i,<t})}$ is the token-level importance ratio,  
$A_{i,t}$ is the token-level advantage, and $(\epsilon_l,\epsilon_h)$ are asymmetric clipping bounds that allow different tolerances for underestimation and overestimation.

These two strategies allow stable and efficient optimization of the DSL agent, with DAPO showing better convergence in practice.

\subsubsection{Reward Function Design}

To optimize the DSL-based retrieval agent, we design a composite reward function that balances structural correctness, execution reliability, semantic accuracy, and response efficiency. The total reward is formulated as
\[
R = S_f + S_e + S_r + S_l,
\]
where each component provides distinct feedback to guide policy optimization. \textbf{Format Reward} (\(S_f\)) encourages syntactic compliance by rewarding outputs that correctly encapsulate reasoning and retrieval logic within predefined DSL tags (e.g., \texttt{<query>...</query>}).  
\textbf{Execution Reward} (\(S_e\)) measures the executability of the generated DSL statement, assigning higher rewards to programs that can be successfully parsed and executed by the retrieval backend.  
\textbf{Result Reward} (\(S_r\)) evaluates the alignment between retrieved results and the reference set, computed as a function of retrieval precision and recall.  
\textbf{Length Reward} (\(S_l\)) penalizes overly long outputs to promote concise and latency-efficient responses, ensuring the agent maintains high utility under real-world inference.

\section{Experiments}

\begin{table*}[ht]
\centering
\caption{
Retrieval performance and latency (ms) comparison on Email and ArxivQA datasets}
\resizebox{1\textwidth}{!}{
\begin{tabular}{l|l|cccc|cc}
\hline
\multirow{2}{*}{\textbf{Category}} & \multirow{2}{*}{\textbf{Model}} & \multicolumn{4}{c|}{\textbf{Email Dataset}} & \multicolumn{2}{c}{\textbf{ArxivQA}} \\
\cline{3-8}
 & & \textbf{Hit@1} & \textbf{Hit@3} & \textbf{MRR} & \textbf{Lat.} & \textbf{NDCG@5} & \textbf{Lat.} \\
\hline
\multirow{2}{*}{\begin{tabular}[c]{@{}l@{}}Traditional\\ Retrieval\end{tabular}} 
 & BM25 & 42.1 & 64.7 & 53.2 & 18 & 31.6 & 25 \\
 & ColQwen & 55.3 & 78.9 & 65.4 & 35 & 73.9 & 42 \\
\hline
\multirow{3}{*}{LLM Based} 
 & Qwen3-4B & 58.3 & 80.1 & 69.2 & 95 & 55.4 & 105 \\
 & Qwen3-8B & 62.1 & 83.6 & 72.5 & 140 & 61.2 & 155 \\
 & GPT-4o & 68.9 & 88.7 & 78.8 & \textbf{30} & 70.5 & \textbf{35} \\
\hline
\multirow{4}{*}{DSL-R1} 
 & Qwen3-4B + GRPO & 69.5 & 88.9 & 78.6 & 55 & 74.8 & 62 \\
 & Qwen3-4B + DAPO & 71.7 & 90.6 & 80.9 & 52 & 75.1 & 58 \\
 & Qwen3-8B + GRPO & 75.8 & 92.9 & 84.6 & 82 & 77.4 & 90 \\
 & Qwen3-8B + DAPO & \textbf{77.9}$^{*}$ & \textbf{94.2}$^{*}$ & \textbf{86.1} & 78 & \textbf{81.2}$^{*}$ & 85 \\
\hline
\end{tabular}
}
\label{tab:email_main_results}
\end{table*}

\subsection{Setup}
\textbf{Datasets.} We evaluate on our industrial Email dataset and ArxivQA~\cite{faysse2024colpaliefficientdocumentretrieval} benchmarks. Both tasks require handling complex interactions between structured and unstructured data. Each query pairs a user request with heterogeneous candidates. Please refer to the Appendix~\ref{datasets} for specific data formats.

\textbf{Baselines.} We compare four groups: 
(1) \textit{Traditional retrieval models} such as BM25~\citep{robertson2009probabilistic} and ColQwen~\citep{colqwen2024}, 
(2) \textit{LLM baselines} (Qwen3-4B, Qwen3-8B, GPT-4o~\citep{openai2024gpt4o}) without RL optimization, 
and (4) our \textit{DSL-R1 agents} trained with GRPO~\citep{grpo2024} and DAPO~\citep{dapo2025}.

\textbf{Metrics.} 
We report standard retrieval metrics: Hit@k, Mean Reciprocal Rank (MRR), and nDCG, 
which measure the ranking quality of generated DSL queries in our setting. 
Latency (ms/query) is also included to reflect runtime efficiency. 
Formal definitions and equations for all metrics are provided in~Appendix~\ref{metrics}.

\subsection{Main Results}
Table~\ref{tab:email_main_results} summarizes the overall results. 
Our DSL-R1 agents outperform all baselines across all metrics, confirming the effectiveness of reinforcement learning in aligning the DSL generator with downstream execution quality. 
Compared with pretrained LLMs, RL optimization improves Hit@1 by 9.6–15.8 points and MRR by up to 13.6 points. 
Among the two optimization strategies, DAPO consistently yields higher accuracy and more stable convergence than GRPO on both the 4B and 8B backbones. 
In addition, reinforcement learning substantially reduces response latency by regularizing output length: \textit{Qwen3-8B + DAPO} achieves 78\,ms/query, a 45\% reduction compared with the vanilla pretrained model.

\subsection{Ablation Studies}

Table~\ref{tab:email_ablation_final} quantifies the contribution of each reward component on our Email dataset.  
Removing any term reduces accuracy, confirming their complementary effects.  
Format and execution rewards ($S_f$, $S_e$) are essential for generating valid and executable DSLs, while the result reward ($S_r$) enhances semantic correctness and the length reward ($S_l$) controls verbosity for faster responses.  
These rewards yield a 15.8-point improvement over the pretrained baseline.

Our ablation analysis reveals consistent trends: removing the SQL filter lowers precision by losing attribute constraints, excluding the vector retrieval module hurts recall on paraphrased queries, and omitting the feedback signal slows convergence and weakens cross-domain generalization.

\begin{table}[t]
\centering
\caption{Ablation study of reward components}
\resizebox{1\linewidth}{!}{
\begin{tabular}{lc}
\hline
\textbf{Reward Function} & \textbf{Accuracy (\%)} \\
\hline
Qwen3-8B (pretrained) & 62.1 \\
\hline
$S_f + S_e + S_r + S_l$ & \textbf{77.9} \\
\quad - w/o $S_f$ (Format Reward) & 75.3 \, (\,$\downarrow$\,2.6) \\
\quad - w/o $S_e$ (Execution Reward) & 75.6 \, (\,$\downarrow$\,2.3) \\
\quad - w/o $S_r$ (Result Reward) & 77.0 \, (\,$\downarrow$\,0.9) \\
\quad - w/o $S_l$ (Length Reward) & 75.9 \, (\,$\downarrow$\,2.0) \\
\hline
\end{tabular}
}
\label{tab:email_ablation_final}
\end{table}

\section{Conclusion}
We introduced \model, unifying SQL logic and vector retrieval in a single DSL, optimized via reward-guided RL. To facilitate rigorous evaluation, we constructed a benchmark encompassing both structured and unstructured retrieval tasks. Across these scenarios, \model\ demonstrates superior accuracy and robustness.

\section{Limitations}

While our DSL-R1 framework demonstrates strong retrieval performance and generality, several limitations remain.  
First, the current system operates in a single-turn setting and does not yet support multi-turn query refinement or conversational retrieval, which would require temporal context modeling and dialogue-state tracking.  
Second, the framework currently adopts a single-agent reinforcement process; extending it to a multi-agent or collaborative setting, such as coordinated query generation, execution, and verification, could further enhance interpretability and robustness.  
Third, although our DSL syntax unifies symbolic and vector-based retrieval, the present experiments are limited to textual and structured inputs.  
Extending the framework to handle  inputs including images, audio, and videos, as well as cross-modal reasoning, remains an open direction.  
Finally, our rule-based reward design simplifies supervision but may constrain adaptability when scaling to more complex and heterogeneous real-world datasets.

\bibliography{custom}

\appendix
\clearpage

\section{}
\subsection{Datasets}
\label{datasets}
We evaluate the alignment of structured metadata with unstructured content across two distinct domains. For each scenario, we utilize 2,000 samples for training and a separate 500 samples for evaluation.

\noindent\textbf{Industrial Email.} This in-house dataset represents enterprise workflows with heterogeneous attachments (e.g., invoices, images). The detailed keyword taxonomy and schema definitions are presented in Table~\ref{tab:schema_def}. We adopted the STaRK framework~\cite{wu24stark} to generate high-quality retrieval instances. For each query, we first select a gold document, then construct a challenging candidate pool by sampling negatives based on structured metadata overlap (Stage 1) and unstructured content similarity (Stage 2). Finally, composite queries are synthesized using the gold document's attributes. We utilize 2,000 samples for training and a distinct 500 samples for evaluation. As shown in Table~\ref{tab:email_stats_refined}, our dataset features complex thread structures and a balanced distribution of structure-dominated versus content-dominated queries.

\begin{table}[h]
    \centering
    \caption{Statistical profile of the Industrial Email dataset}
    \label{tab:email_stats_refined}
    
    \resizebox{\linewidth}{!}{%
    \begin{tabular}{lrrr} 
        \toprule
        \textbf{Metric} & \textbf{Train} & \textbf{Eval} & \textbf{Total} \\
        \midrule
        
        \multicolumn{4}{l}{\textsc{\textbf{Corpus Topology}}} \\
        \quad Corpus Size ($N$) & 2,000 & 500 & 2,500 \\
        \quad Avg. Doc. Length (tokens) & 154.2 & 156.8 & 154.7 \\
        
        \midrule
        
        \multicolumn{4}{l}{\textsc{\textbf{Query Constraint Statistics}}} \\
        \textit{Avg. Keyword Counts ($k$)} & & & \\
        \quad Structured Keys ($k_{str}$) & 1.8 & 2.1 & 2.0 \\
        \quad Unstructured Keys ($k_{uns}$) & 3.1 & 3.0 & 3.1 \\
        \quad \textit{Total Complexity ($k_{total}$)} & 4.9 & 5.1 & 5.0 \\

        \cmidrule(lr){2-4} 
        
        \textit{Modality Distribution (\%)} & & & \\
        \quad \textbf{Structure-Dominated} ($k_{str} > k_{uns}$) & 30.0 & 28.5 & 29.7 \\
        \quad \textbf{Content-Dominated} ($k_{uns} > k_{str}$) & 45.0 & 46.5 & 45.3 \\
        \quad \textbf{Balanced} ($k_{str} \approx k_{uns}$) & 25.0 & 25.0 & 25.0 \\
        
        \bottomrule
    \end{tabular}%
    }
\end{table}

\begin{table}[h]
    \centering
    \caption{Definition of keywords in the Industrial Email dataset, categorized by information type}
    \label{tab:schema_def}
    \resizebox{\linewidth}{!}{%
    \begin{tabular}{llp{6cm}}
        \toprule
        \textbf{Category} & \textbf{Keyword} & \textbf{Description} \\
        \midrule
        \multirow{8}{*}{\textbf{Structured Info}} 
         & \texttt{account\_email} & The email address of the account owner. \\
         & \texttt{received\_date} & Timestamp when the email was received. \\
         & \texttt{is\_draft} & Boolean flag indicating if the email is a draft. \\
         & \texttt{draft\_created\_date} & Timestamp for draft creation. \\
         & \texttt{draft\_modified\_date} & Timestamp for last draft modification. \\
         & \texttt{is\_read} & Boolean status indicating if the email has been read. \\
         & \texttt{is\_starred} & Boolean status indicating if the email is starred. \\
         & \texttt{is\_archived} & Boolean status indicating if the email is archived. \\
         & \texttt{thread\_msg\_count} & Number of messages in the current thread. \\
        \midrule
        \multirow{9}{*}{\textbf{Unstructured Info}} 
         & \texttt{sender\_email} & Email address of the sender. \\
         & \texttt{sender\_name} & Display name of the sender. \\
         & \texttt{recipient\_list} & List of primary recipient email addresses. \\
         & \texttt{cc\_list} & List of carbon copy (CC) recipients. \\
         & \texttt{bcc\_list} & List of blind carbon copy (BCC) recipients. \\
         & \texttt{folder\_labels} & Labels or folders associated with the email. \\
         & \texttt{attachment\_list} & Metadata of files attached to the email. \\
         & \texttt{subject} & The subject line text of the email. \\
         & \texttt{content} & The main body text of the email message. \\
        \bottomrule
    \end{tabular}%
    }
\end{table}

\noindent\textbf{Scientific Papers.} This domain focuses on retrieving information from visually rich documents characterized by complex layouts containing embedded images, tables, and formulas, enriched with structured metadata such as domain, category, author, and publish time. For evaluation, we strictly employ the ArXivQA subset from the ViDoRe benchmark~\cite{faysse2024colpaliefficientdocumentretrieval} to ensure comparability. Conversely, training utilizes the distinct original ArXivQA dataset~\cite{li-etal-2024-multimodal-arxiv}, ensuring the model learns generalizable alignment features without overfitting to the benchmark distribution.


\subsection{Metrics}
\label{metrics}
We evaluate system performance from both retrieval accuracy and program executability. Let $\mathcal{Q}$ denote the set of evaluation queries, and for each query $q \in \mathcal{Q}$, let $r_q$ be the rank position of the first relevant (ground-truth) item in the retrieved list.

\paragraph{Hit@k.}
Hit@k measures whether the correct item appears within the top-$k$ retrieved results:
\begin{equation}
\mathrm{Hit@}k = \frac{1}{|\mathcal{Q}|} \sum_{q \in \mathcal{Q}} \mathbf{1}\!\left[r_q \le k\right],
\end{equation}
where $\mathbf{1}[\cdot]$ is the indicator function.

\paragraph{MRR.}
MRR evaluates ranking quality by assigning higher scores to correct items appearing earlier in the ranking:
\begin{equation}
\mathrm{MRR} = \frac{1}{|\mathcal{Q}|} \sum_{q \in \mathcal{Q}} \frac{1}{r_q},
\end{equation}

\paragraph{nDCG@5.}
nDCG@5 accounts for graded relevance and penalizes lower-ranked correct items using logarithmic discounting. The discounted cumulative gain at cutoff $K=5$ is defined as:
\begin{equation}
\mathrm{DCG@}5 = \sum_{i=1}^{5} \frac{2^{\mathrm{rel}_i} - 1}{\log_2(i+1)},
\end{equation}
where $\mathrm{rel}_i$ denotes the relevance score of the item at rank $i$. The normalized score is obtained by dividing by the ideal DCG:
\begin{equation}
\mathrm{nDCG@}5 = \frac{\mathrm{DCG@}5}{\mathrm{IDCG@}5},
\end{equation}


\paragraph{Latency.}
We additionally report average latency, measured in milliseconds per query:
\begin{equation}
\mathrm{Latency} = \frac{1}{|\mathcal{Q}|} \sum_{q \in \mathcal{Q}} t_q,
\end{equation}
where $t_q$ is the wall-clock execution time for query $q$, including both retrieval and DSL generation.

\subsection{Implementation Details.}
We implemented our training pipeline utilizing the \texttt{verl} framework~\cite{sheng2024hybridflow}, adopting the ColQwen2.5~\cite{colqwen2024} model as our RAG embedding backbone. Fine-tuning was conducted on a cluster of 4 NVIDIA H100 GPUs for 10 epochs with a learning rate of $1 \times 10^{-6}$. All inference benchmarks and latency measurements were performed on a single NVIDIA A100 GPU.

\subsection{Query-DSL Cases}
\label{cases}

The following examples demonstrate the mapping between natural language user queries and our Domain Specific Language (DSL). Each case illustrates how a user's intent is parsed into a hybrid structure consisting of an SQL statement for relational filtering and a vector query list for semantic retrieval.

\begin{description}
    \item[Case 1: Budget Drafts] \hfill \\
    \textbf{Query:} ``Find starred draft emails modified in the last 7 days containing `budget' in the body.''
    
    \textbf{DSL Output:}
    \begin{lstlisting}[language=SQL]
{
  "sql": "SELECT DISTINCT message_id FROM email WHERE is_draft = 1 AND draft_modified_date >= date('now', '-7 day') AND is_starred = 1 AND message_id IN <vector_0>",
  "vector_query_list": [
    {"email_content": "budget"}
  ]
}
    \end{lstlisting}

    \item[Case 2: Important Folders] \hfill \\
    \textbf{Query:} ``Find folders where the label name contains `important'.''
    
    \textbf{DSL Output:}
    \begin{lstlisting}[language=SQL]
{
  "sql": "SELECT DISTINCT json_extract(json_each.value, '$.id') AS folder_id FROM email, json_each(email.folder_labels) WHERE json_extract(json_each.value, '$.id') IN <vector_0>",
  "vector_query_list": [
    {"folder_labels": "important"}
  ]
}
    \end{lstlisting}

    \item[Case 3: Contract Attachments] \hfill \\
    \textbf{Query:} ``Find attachment IDs for emails where the subject contains `contract'.''
    
    \textbf{DSL Output:}
    \begin{lstlisting}[language=SQL]
{
  "sql": "SELECT DISTINCT json_extract(json_each.value, '$.id') AS attachment_id FROM email, json_each(email.attachment_list) WHERE message_id IN <vector_0> AND json_extract(json_each.value, '$.id') IN <vector_1>",
  "vector_query_list": [
    {"subject": "contract"},
    {"attachment_list": "contract"}
  ]
}
    \end{lstlisting}
\end{description}
\end{document}